%% file: main.tex
\documentclass[conference]{IEEEtran}
\IEEEoverridecommandlockouts

\usepackage{cite}
\usepackage{amsmath,amssymb,amsfonts}
\usepackage{graphicx}
\usepackage{algorithm}
\usepackage{algpseudocode}
\usepackage{qcircuit}
\usepackage[caption=false,font=footnotesize]{subfig}
\usepackage{tikz}

\def\BibTeX{{\rm B\kern-.05em{\sc i\kern-.025em b}\kern-.08em
    T\kern-.1667em\lower.7ex\hbox{E}\kern-.125emX}}
\begin{document}

\title{Efficient Time-Aware Partitioning of Quantum Circuits for Distributed Quantum Computing\\
\thanks{This research was supported by Deakin University and CSIRO's Data61 through the Next Generation Quantum Graduate Program (118-NGQGP).}
}

\author{\IEEEauthorblockN{Raymond P. H. Wu}
\IEEEauthorblockA{\textit{School of Information Technology} \\
\textit{Deakin University}\\
Burwood, Victoria 3125, Australia \\
rayphwu@gmail.com}
\and
\IEEEauthorblockN{Chathu Ranaweera}
\IEEEauthorblockA{\textit{School of Information Technology} \\
\textit{Deakin University}\\
Burwood, Victoria 3125, Australia \\
chathu.ranaweera@deakin.edu.au}
\and
\IEEEauthorblockN{Sutharshan Rajasegarar}
\IEEEauthorblockA{\textit{School of Information Technology} \\
\textit{Deakin University}\\
Burwood, Victoria 3125, Australia \\
sutharshan.rajasegarar@deakin.edu.au}
\and
\IEEEauthorblockN{Ria Rushin Joseph}
\IEEEauthorblockA{\textit{School of Information Technology} \\
\textit{Deakin University}\\
Burwood, Victoria 3125, Australia \\
ria.joseph@deakin.edu.au}
\and
\IEEEauthorblockN{Jinho Choi}
\IEEEauthorblockA{\textit{School of Electrical and Mechanical Engineering} \\
\textit{The University of Adelaide}\\
Adelaide, South Australia 5005, Australia \\
jinho.choi@adelaide.edu.au}
\and
\IEEEauthorblockN{Seng W. Loke}
\IEEEauthorblockA{\textit{School of Information Technology} \\
\textit{Deakin University}\\
Burwood, Victoria 3125, Australia \\
seng.loke@deakin.edu.au}
}

\maketitle

\begin{abstract}
    To overcome the physical limitations of scaling monolithic quantum computers, distributed quantum computing (DQC) interconnects multiple smaller-scale quantum processing units (QPUs) to form a quantum network.
    However, this approach introduces a critical challenge, namely the high cost of quantum communication between remote QPUs incurred by quantum state teleportation and quantum gate teleportation.
    To minimize this communication overhead, DQC compilers must strategically partition quantum circuits by mapping logical qubits to distributed physical QPUs.
    Static graph partitioning methods are fundamentally ill-equipped for this task as they ignore execution dynamics and underlying network topology, while metaheuristics require substantial computational runtime.
    In this work, we propose a heuristic based on beam search to solve the circuit partitioning problem.
    Our time-aware algorithm incrementally constructs a low-cost sequence of qubit assignments across successive time steps to minimize overall communication overhead.
    The time and space complexities of the proposed algorithm scale quadratically with the number of qubits and linearly with circuit depth, offering a significant computational speedup over common metaheuristics.
    We demonstrate that our proposed algorithm consistently achieves significantly lower communication costs than static baselines across varying circuit sizes, depths, and network topologies, providing an efficient compilation tool for near-term distributed quantum hardware.
\end{abstract}


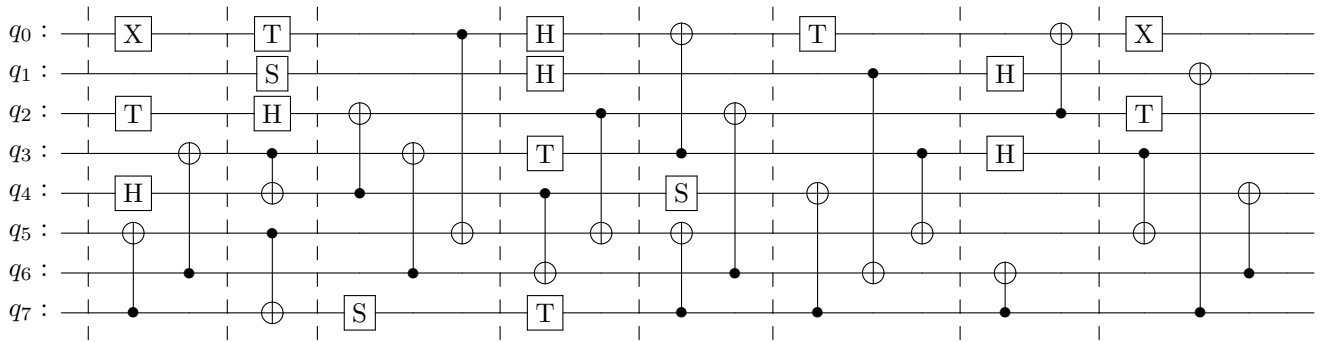
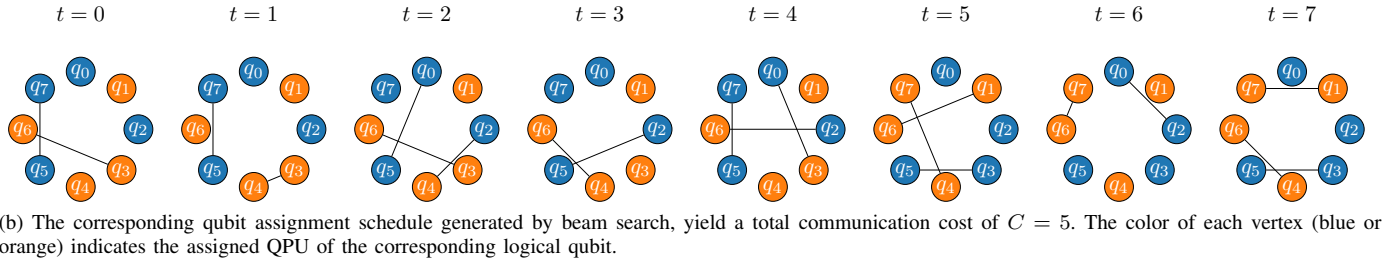
\begin{figure*}[!t]
    \centering
    \subfloat[A random quantum circuit with number of qubits $N=8$ and circuit depth $T=8$.]{
        \resizebox{\textwidth}{!}{\input{fig1a.tex}}
    \label{fig1a}}
    \hfil
    \subfloat[The corresponding qubit assignment schedule generated by beam search, yield a total communication cost of $C=5$.
    The color of each vertex (blue or orange) indicates the assigned QPU of the corresponding logical qubit.]{
        \resizebox{\textwidth}{!}{\input{fig1b.tex}}
    \label{fig1b}}
    \caption{An illustrative example of quantum circuit partitioning for a system with $k=2$ QPUs of equal capacity $c_j=4$.}
    \label{fig1}
\end{figure*}

\section{Introduction}\label{seci}
Existing gate-based quantum computers remain confined to the noisy intermediate-scale quantum (NISQ) regime\cite{preskill2018}, with quantum processing units (QPUs) consisting of thousands of noisy physical qubits.
This capacity is insufficient for practical quantum advantage, which requires scaling to millions of fault-tolerant physical qubits.
To overcome the physical limitations of monolithic scaling, distributed quantum computing (DQC) interconnects multiple smaller-scale QPUs, forming a quantum network to collectively execute computational tasks that exceed the capacity of any single device\cite{loke2023,caleffi2024,barral2025}.

However, quantum communication between remote QPUs relies on Einstein–Podolsky–Rosen (EPR) pairs, whose generation and distribution are significantly more costly and fragile to decoherence than local operations\cite{nielsen2010}.
DQC relies on two fundamental communication protocols.
Specifically, quantum state teleportation reconstructs an unknown quantum state at a remote destination without physically transporting the qubit at the source, whereas quantum gate teleportation executes a multi-qubit gate, such as a controlled NOT (CNOT) gate, between physical qubits located on distinct QPUs.
Because both protocols consume identical resources (one EPR pair and two classical bits), DQC compilers must strategically map logical qubits to distributed physical QPUs to minimize communication overhead\cite{ferrari2021,ferrari2023,cuomo2023}.
This process known as quantum circuit partitioning transforms the task of executing a monolithic quantum circuit into a set of smaller sub-circuits that can be executed concurrently across multiple interconnected QPUs.
However, finding the optimal circuit partitioning strategy depends entirely on the specific quantum circuit, hardware constraints, and underlying network topology.
A quantum circuit can be abstracted as a temporal sequence of graphs, where edges represent two-qubit operations.
Even if this temporal dimension is collapsed into a single static graph, the simplified circuit partitioning problem reduces to the $k$-way graph partitioning problem, which is NP-hard\cite{mao2023}.
Consequently, the temporal formulation inherits this NP-hardness while introducing a dimension of combinatorial complexity, exponentially expanding the solution space.
Because obtaining an exact optimal qubit assignment schedule is intractable for practical-scale quantum circuits, DQC compilers must rely on approximation algorithms and heuristics.

Standard graph partitioning algorithms, most notably METIS\cite{karypis1998}, use multilevel partitioning algorithms to compute a partition at the coarsest graph and then refine the solution.
These standard methods have been applied to the quantum circuit partitioning problem\cite{daei2020}.
Hypergraph partitioning extend this by representing groups of two-qubit gates as hyperedge\cite{andres2019,cambiucci2023}.
However, these static approaches are fundamentally ill-equipped as they ignore execution dynamics, produce a single, fixed qubit assignment for the entire duration of the circuit, and cannot model the underlying network topology.
Time-aware methods incorporate the temporal dimension of a quantum circuit by producing a sequence of qubit assignments across discrete time steps\cite{baker2020,burt2024}.
This dynamic mapping allows quantum states of logical qubits to be transferred across distinct QPUs as their multi-qubit gate operands change, minimizing overall communication overhead.
To generate qubit assignment schedules, existing literature relies heavily on metaheuristics, particularly evolutionary algorithms\cite{houshmand2020,sunkel2024,sunkel2025} and simulated annealing\cite{mao2023,sunkel2025}, to explore the vast solution space.
These metaheuristics demonstrate significant reductions in communication cost compared to static graph partitioning baselines at the expense of computational runtime.
Although their objective functions integrate the underlying network topology, hardware constraints are imposed via penalties, which is computationally inefficient.
Collectively, the existing literature presents a critical trade-off where fast static graph partitioning methods yield suboptimal solutions, while dynamic metaheuristics require substantial computational runtime.

To minimize communication overhead in DQC without incurring prohibitive computational runtime, we propose a time-aware heuristic based on beam search\cite{lowerre1976} to solve the circuit partitioning problem.
Beam search limits the search breadth at each layer of the search tree by the beam width $\beta$, providing a tunable trade-off between computational tractability and exhaustive exploration.
Our algorithm incrementally constructs a low-cost sequence of qubit assignments $\mathcal{S}^*=(S_0,S_1,\dots,S_{T-1})$ across successive time steps over the circuit depth $T$ to minimize the objective function $C(\mathcal{S})$.
If the search parameters scale linearly with number of qubits $N$, the time and space complexities of the proposed algorithm both scale as $O(N^2T\beta)$, offering a significant computational speedup.
We demonstrate that our proposed time-aware heuristic consistently finds solutions with lower communication cost than static graph partitioning algorithm baselines like METIS and has lower complexity than common metaheuristics, providing an efficient tool for DQC compilation.

\section{Circuit Partitioning Problem}\label{secii}
\subsection{Graph Representation of Quantum Circuits}\label{seciia}
A quantum circuit acting on a set of $N$ logical qubits, $V=\{q_0,q_1,\dots,q_{N-1}\}$, is a sequence of operations applied across discrete time steps over the circuit depth $T$.
We represent a quantum circuit as a sequence of $T$ undirected graphs, $\mathcal{G}=(G_0,G_1,\dots,G_{T-1})$.
Each graph $G_t=(V,E_t)$, corresponds to the operations applied on the logical qubits at time step $t$.
The set of vertices $V$ is constant for all graphs and represents the logical qubits.
The set of edges $E_t$ contains an edge $(q_i,q_j)$ if and only if a two-qubit gate is applied between logical qubits $q_i$ and $q_j$ at time step $t$.
Single-qubit gates do not form edges, as they do not require quantum communication.
This representation encodes the execution dynamics of the circuit.

\subsection{Problem Formulation}\label{seciib}
Given a system of $k$ QPUs, $P=\{n_0,n_1,\dots,n_{k-1}\}$, where each QPU $n_j$ has a maximum capacity of $c_j$ qubits, our goal is to find an optimal mapping for each logical qubit at each time step to minimize the total communication cost.
A qubit assignment schedule, $\mathcal{S}=(S_0,S_1,\dots,S_{T-1})$, is a sequence of $T$ assignment functions.
Each function $S_t:V\to P$ maps every logical qubit in the set $V$ to a specific physical QPU in the set $P$ at a given time step $t$.
The schedule $\mathcal{S}$ must satisfy the hardware capacity constraints of the QPUs at all times.
That is, for every time step $t\in\{0,\dots,T-1\}$ and for every QPU $n_j\in P$, the number of logical qubits assigned to $n_j$ must not exceed its capacity $c_j$:
\begin{equation}
\left|\{q_i\in V:S_t(q_i)=n_j\}\right|\le c_j.\label{eqiib1}
\end{equation}

The total communication cost $C$ of a given assignment $\mathcal{S}$ is composed of two components, namely the cost of quantum state teleportation $C_{\text{state}}$ and the cost of quantum gate teleportation $C_{\text{gate}}$.
Let $D$ be the $k \times k$ QPU distance matrix, where $D(n_i,n_j)$ is the shortest path distance from QPU $n_i$ to QPU $n_j$.
The cost of executing quantum state teleportation and quantum gate teleportation depends on the distance between the source and destination QPUs.
A quantum state teleportation is required whenever the assigned QPU of a logical qubit changes between consecutive time steps (i.e., $S_{t-1}(q_i)\neq S_t(q_i)$).
For a given schedule $\mathcal{S}=(S_0,\dots,S_{T-1})$, the cost of state teleportation $C_{\text{state}}$ is the sum of costs for all logical qubits that change QPUs between consecutive time steps:
\begin{equation}
C_{\text{state}}(\mathcal{S})=\sum_{t=1}^{T-1}\sum_{q_i\in V}D(S_{t-1}(q_i),S_t(q_i)).\label{eqiib2}
\end{equation}
A quantum gate teleportation is required at time step $t$ for every two-qubit gate $(q_i,q_j)\in E_t$ where the qubits are assigned to distinct QPUs (i.e., $S_t(q_i)\neq S_t(q_j)$).
The cost of gate teleportation $C_{\text{gate}}$ is the sum of costs for all two-qubit gates at each time step $t$ that are split across distinct QPUs:
\begin{equation}
C_{\text{gate}}(\mathcal{S})=\sum_{t=0}^{T-1}\sum_{(q_i, q_j)\in E_t}D(S_t(q_i),S_t(q_j)).\label{eqiib3}
\end{equation}
Let $w_{\text{state}}$ and $w_{\text{gate}}$ be the weighting factors for quantum state teleportation and quantum gate teleportation, respectively.
These weights are parameters that reflect the relative physical cost of the two operations, such as latency and infidelity.
The total cost function $C(\mathcal{S})$ is the weighted sum:
\begin{equation}
    C(\mathcal{S})=w_{\text{state}}C_{\text{state}}(\mathcal{S})+w_{\text{gate}}C_{\text{gate}}(\mathcal{S}).\label{eqiib4}
\end{equation}
The goal of the circuit partitioning problem is to find an optimal schedule $\mathcal{S}^*$ that minimizes this total cost function
\begin{equation}
    \mathcal{S}^*=\arg\min_{\mathcal{S}}C(\mathcal{S}),\label{eqiib5}
\end{equation}
while satisfying the capacity constraints in \eqref{eqiib1} for all QPUs.

The circuit partitioning problem inherits the NP-hardness of the $k$-way graph partitioning problem of a static weighted graph.
The number of ways to assign $N$ qubits into $k$ QPUs, while satisfying the capacity constraints $c_j$, over the circuit depth $T$, scales as $O(k^{NT})$.
This vast solution space makes finding an exact solution computationally intractable, necessitating the use of approximation algorithms and heuristics.

\section{Beam Search Algorithm for Quantum Circuit Partitioning}\label{seciii}
We propose a time-aware heuristic based on beam search to solve the circuit partitioning problem (Algorithm~\ref{algorithm1}).
Beam search limits the search breadth at each layer of the search tree by the beam width $\beta$, providing a tunable trade-off between computational tractability and exhaustive exploration.
Our algorithm incrementally constructs a low-cost sequence of qubit assignments $\mathcal{S}^*=(S_0,S_1,\dots,S_{T-1})$ across successive time steps over the circuit depth $T$ to minimize the objective function $C(\mathcal{S})$.

The search initializes at time step $t=0$ with a beam of $\beta$ randomly generated qubit assignments that satisfy the capacity constraints $c_j$.
At each subsequent time step $t$, the algorithm expands the search tree by generating $\Gamma$ candidate assignments for each of the $\beta$ partial schedules using four distinct strategies:
\begin{description}[\IEEEsetlabelwidth{Diversification:}]
    \item[Preservation:] Retains the previous assignment $S_{t-1}$.
    \item[Mitigation:] Identifies all two-qubit gates $(q_i,q_j)\in E_t$ where the logical qubits are assigned to distinct physical QPUs.
    For each such edge, it generates two new candidate assignments: one by assigning $q_i$ to the QPU of $q_j$, and another by assigning $q_j$ to the QPU of $q_i$.
    \item[Swaps:] Generates $\Gamma_\text{swaps}$ candidate assignments by randomly swapping the physical QPU assignments of two distinct logical qubits.
    \item[Diversification:] Generates $\Gamma_\text{random}$ valid random candidate assignments.
\end{description}
Any candidate assignment that violates the hardware capacity constraints $c_j$ is immediately discarded.
The algorithm then calculates the cumulative communication cost for the partial schedules using \eqref{eqiib4}.
These schedules are sorted by cost, and only the top $\beta$ schedules are retained to form the new beam for the next time step, pruning the remainder of the search space.
This process continues until the last time step $T-1$ is evaluated.
The complete qubit assignment schedule $\mathcal{S}^*$ with the lowest total communication cost is then selected as the solution.

The time and space complexities of the proposed algorithm scale as $O(NT\beta\Gamma)$.
If the search parameters scale linearly with the number of qubits $N$, their upper bound scales as $O(N^2T\beta)$, offering a significant computational speedup.

\begin{algorithm}
\caption{Beam Search for Quantum Circuit Partitioning}
\label{algorithm1}
\begin{algorithmic}[1]
\Require Quantum circuit $\mathcal{G}$, distance matrix $D$, beam width $\beta$, QPU capacities $\{c_j\}$, weights $w_{\text{state}}, w_{\text{gate}}$, expansion parameters $\Gamma$
\Ensure Best found qubit assignment schedule $\mathcal{S}^*$
\State $\mathcal{B}_{0}\gets$ \Call{Initialize}{$\beta,\{c_j\}$}
\For{$t=1$ to $T-1$}
    \State $\mathcal{L}_t\gets\emptyset$
    \ForAll{$\mathcal{S}_{t-1}\in\mathcal{B}_{t-1}$}
        \ForAll{$S_t\in$ \Call{Candidates}{$\mathcal{S}_{t-1},G_t,\Gamma$}}
            \If{\Call{Valid}{$S_t,\{c_j\}$}}
                \State $\mathcal{S}_t\gets\mathcal{S}_{t-1}\oplus S_t$
                \State $C(\mathcal{S}_t)\gets C(\mathcal{S}_{t-1})+$ \Call{Cost}{$\mathcal{S}_{t-1}, S_t, G_t$}
                \State $\mathcal{L}_t\gets\mathcal{L}_t\cup\{\mathcal{S}_t\}$
            \EndIf
        \EndFor
    \EndFor
    \State $\mathcal{B}_t\gets$ \Call{Top}{$\mathcal{L}_t,C,\beta$}
\EndFor
\State \Return $\arg\min_{\mathcal{S}\in\mathcal{B}_{T-1}}C(\mathcal{S})$
\Statex
\Function{Candidates}{$\mathcal{S}_{t-1},G_t,\Gamma$}
    \State $\mathcal{A}\gets$ \Call{Preservation}{$S_{t-1}$}
    \State $\mathcal{A}\gets\mathcal{A}\cup$ \Call{Mitigation}{$S_{t-1},E_t$}
    \State $\mathcal{A}\gets\mathcal{A}\cup$ \Call{Swaps}{$S_{t-1},\Gamma_\text{swaps}$}
    \State $\mathcal{A}\gets\mathcal{A}\cup$ \Call{Diversification}{$\Gamma_\text{random}$}
    \State \Return $\mathcal{A}$
\EndFunction
\Function{Cost}{$\mathcal{S}_{t-1},S_t,G_t$}
    \State $\Delta C_{\text{state}}\gets\sum_{q_i\in V}D(S_{t-1}(q_i),S_t(q_i))$
    \State $\Delta C_{\text{gate}}\gets\sum_{(q_i,q_j)\in E_t}D(S_t(q_i),S_t(q_j))$
    \State \Return $w_{\text{state}}\Delta C_{\text{state}}+w_{\text{gate}}\Delta C_{\text{gate}}$
\EndFunction
\end{algorithmic}
\end{algorithm}

\section{Results}\label{seciv}
\subsection{Experimental Setup}\label{seciva}
We evaluated our proposed algorithm against METIS, a standard graph-partitioning algorithm, using randomly generated quantum circuits.
At each time step, a gate is applied to a pair of qubits with a probability $0.5$ for being a CNOT gate and $0.5$ for being a single-qubit gate or idle operation.
The beam search algorithm takes the temporal sequence of graphs $\mathcal{G}=(G_0,G_1,\dots,G_{T-1})$ as input.
Conversely, METIS requires a static input.
Therefore, for METIS, we aggregated this sequence into a single weighted graph where the weight $w_E$ of an edge $E(u,v)$ corresponds to the total number of CNOT operations between qubits $q_u$ and $q_v$ across the entire circuit.
We benchmarked $10$ independent random circuit samples and $3$ independent runs for each sample.
For simplicity, we set $w_{\text{state}}/w_{\text{gate}}=1$.
To ensure high solution quality at the expense of increased computational runtime, our experimental analysis determined the search parameter thresholds to be $\beta\ge8N$, $\Gamma_{\text{swaps}}\ge4N$, and $\Gamma_{\text{random}}\ge2N$.

\subsection{Baseline Comparison and Scalability}\label{secivb}

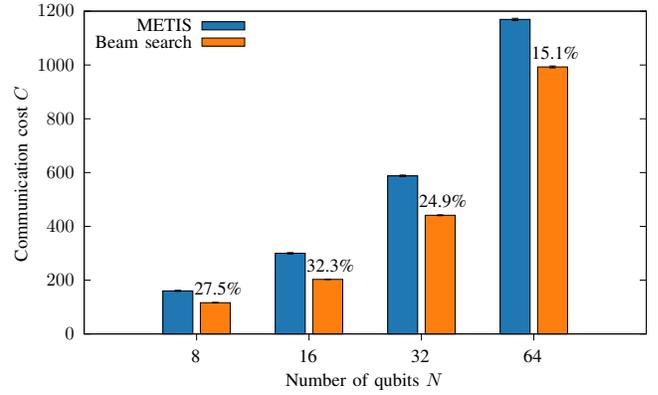
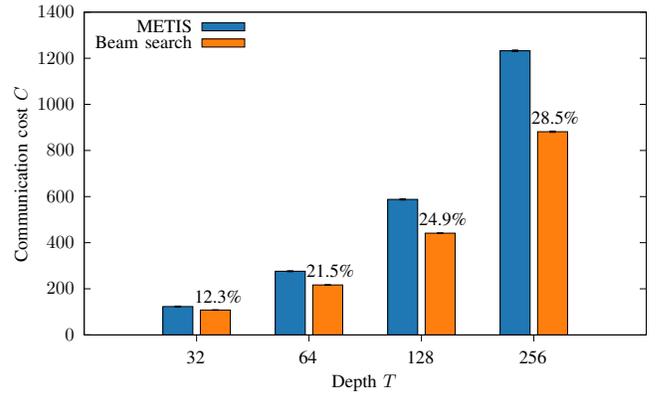
\begin{figure}[!t]
    \centering
    \subfloat[Random quantum circuits with varying logical qubits $N$ and a fixed depth $T=128$.]{
        \resizebox{\columnwidth}{!}{\input{fig2a.tex}}
    \label{fig2a}}
    \hfil
    \subfloat[Random quantum circuits with a fixed logical qubit count $N=32$ and varying depth $T$.]{
        \resizebox{\columnwidth}{!}{\input{fig2b.tex}}
    \label{fig2b}}
    \caption{Comparison of communication costs between METIS and beam search for quantum circuits of various sizes.
    Error bars represent one standard error.}
    \label{fig2}
\end{figure}

We analyze the baseline solution quality for systems with $k=2$ QPUs of equal capacity $c_j=N/2$.
An illustrative example is shown in Fig.~\ref{fig1} for a random quantum circuit with $N=8$ and $T=8$, and its corresponding qubit assignment schedule found by beam search.
Scaling to larger circuits, as shown in Fig.~\ref{fig2}, the beam search algorithm consistently achieves significantly lower communication costs than METIS.
While this cost reduction remains when scaling the number of qubits from $N=8$ to $N=64$ with a fixed circuit depth $T=128$ (Fig.~\ref{fig2a}), the advantage in solution quality widens as the circuit depth increases from $T=32$ to $T=256$ with a fixed number of qubits $N=32$ (Fig.~\ref{fig2b}).
METIS, which finds a static qubit assignment from the weighted graph, is fundamentally ill-equipped to handle the dynamic nature of circuit execution because it optimizes solely for edge cuts and ignores state teleportation as a mechanism to minimize overall costs.
By contrast, our algorithm operates directly on the temporal sequence of graphs, dynamically optimizing the trade-off between state and gate teleportation at each time step.

\subsection{Adaptability to Network Topologies}\label{secivc}

\begin{figure}[!t]
    \centering
    \resizebox{\columnwidth}{!}{\input{fig3.tex}}
    \caption{Comparison of communication costs between METIS and beam search across different network topologies for random quantum circuits with number of qubits $N=32$ and circuit depth $T=128$.
    Error bars represent one standard error.}
    \label{fig3}
\end{figure}
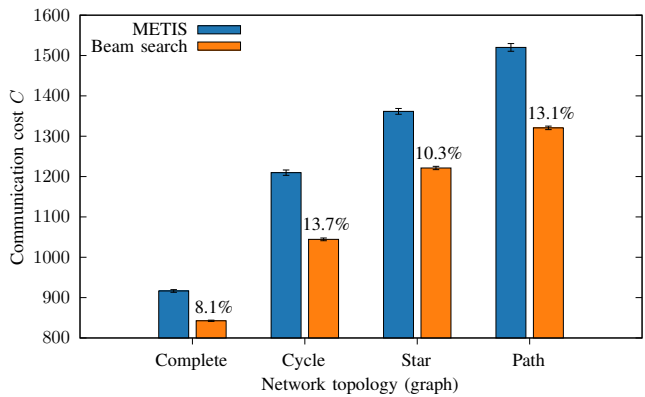

A critical limitation of static graph-partitioning algorithms like METIS is their inability to optimize for the varying communication overhead inherent to a specific physical network topology.
Our proposed beam search algorithm incorporates physical constraints via the distance matrix $D$, penalizing distant state teleportation and gate teleportation.
To demonstrate this advantage, we extended our analysis to systems with $k=4$ QPUs of equal capacity $c_j=N/4$ operating across four distinct network topologies defined by the complete $K_4$, cycle $C_4$, star $K_{1,3}$, and path $P_4$ graphs. 
As shown in Fig.~\ref{fig3}, our algorithm found significantly lower-cost partitions than METIS across all topologies for circuits with number of qubits $N=32$ and circuit depth $T=128$.
Being topology-agnostic, METIS produces the exact same static partition for all four cases, resulting in highly suboptimal assignments for constrained networks.
By contrast, our topology-aware algorithm finds a qubit assignment schedule that respects the underlying physical network.
This proves our algorithm's utility for optimizing the partitioning of quantum circuits on realistic, near-term distributed quantum hardware.

\section{Conclusion}\label{secv}
In this work, we proposed a time-aware heuristic based on beam search to solve the circuit partitioning problem for distributed quantum computing to minimize communication overhead.
Unlike static graph partitioning baselines that ignore execution dynamics, our algorithm incrementally constructs a sequence of qubit assignments across successive time steps, optimizing the trade-off between quantum state teleportation and quantum gate teleportation.
Experimental results demonstrate that our algorithm consistently achieves significantly lower communication costs than standard baselines like METIS, with the communication cost reduction widening as circuit depth increases.
Furthermore, by modeling the network topology via a distance matrix, our algorithm optimizes schedules across different network topologies, whereas topology-agnostic baselines yield suboptimal assignments.
With time and space complexities scaling quadratically with the number of qubits and linearly with circuit depth, our algorithm offers a significant computational speedup over common metaheuristics, providing an efficient compilation tool for near-term distributed quantum hardware.



\end{document}

%% file: fig1a.tex


\scalebox{1.0}{
\Qcircuit @C=1.0em @R=0.2em @!R { \\
	 	\nghost{{q}_{0} :  } & \lstick{{q}_{0} :  } \barrier[0em]{7} & \qw & \gate{\mathrm{X}} & \qw \barrier[0em]{7} & \qw & \gate{\mathrm{T}} \barrier[0em]{7} & \qw & \qw & \qw & \ctrl{5} \barrier[0em]{7} & \qw & \gate{\mathrm{H}} & \qw \barrier[0em]{7} & \qw & \targ & \qw \barrier[0em]{7} & \qw & \gate{\mathrm{T}} & \qw & \qw \barrier[0em]{7} & \qw & \qw & \targ \barrier[0em]{7} & \qw & \gate{\mathrm{X}} & \qw & \qw & \qw & \qw\\
	 	\nghost{{q}_{1} :  } & \lstick{{q}_{1} :  } & \qw & \qw & \qw & \qw & \gate{\mathrm{S}} & \qw & \qw & \qw & \qw & \qw & \gate{\mathrm{H}} & \qw & \qw & \qw & \qw & \qw & \qw & \ctrl{5} & \qw & \qw & \gate{\mathrm{H}} & \qw & \qw & \qw & \targ & \qw & \qw & \qw\\
	 	\nghost{{q}_{2} :  } & \lstick{{q}_{2} :  } & \qw & \gate{\mathrm{T}} & \qw & \qw & \gate{\mathrm{H}} & \qw & \targ & \qw & \qw & \qw & \qw & \ctrl{3} & \qw & \qw & \targ & \qw & \qw & \qw & \qw & \qw & \qw & \ctrl{-2} & \qw & \gate{\mathrm{T}} & \qw & \qw & \qw & \qw\\
	 	\nghost{{q}_{3} :  } & \lstick{{q}_{3} :  } & \qw & \qw & \targ & \qw & \ctrl{1} & \qw & \qw & \targ & \qw & \qw & \gate{\mathrm{T}} & \qw & \qw & \ctrl{-3} & \qw & \qw & \qw & \qw & \ctrl{2} & \qw & \gate{\mathrm{H}} & \qw & \qw & \ctrl{2} & \qw & \qw & \qw & \qw\\
	 	\nghost{{q}_{4} :  } & \lstick{{q}_{4} :  } & \qw & \gate{\mathrm{H}} & \qw & \qw & \targ & \qw & \ctrl{-2} & \qw & \qw & \qw & \ctrl{2} & \qw & \qw & \gate{\mathrm{S}} & \qw & \qw & \targ & \qw & \qw & \qw & \qw & \qw & \qw & \qw & \qw & \targ & \qw & \qw\\
	 	\nghost{{q}_{5} :  } & \lstick{{q}_{5} :  } & \qw & \targ & \qw & \qw & \ctrl{2} & \qw & \qw & \qw & \targ & \qw & \qw & \targ & \qw & \targ & \qw & \qw & \qw & \qw & \targ & \qw & \qw & \qw & \qw & \targ & \qw & \qw & \qw & \qw\\
	 	\nghost{{q}_{6} :  } & \lstick{{q}_{6} :  } & \qw & \qw & \ctrl{-3} & \qw & \qw & \qw & \qw & \ctrl{-3} & \qw & \qw & \targ & \qw & \qw & \qw & \ctrl{-4} & \qw & \qw & \targ & \qw & \qw & \targ & \qw & \qw & \qw & \qw & \ctrl{-2} & \qw & \qw\\
	 	\nghost{{q}_{7} :  } & \lstick{{q}_{7} :  } & \qw & \ctrl{-2} & \qw & \qw & \targ & \qw & \gate{\mathrm{S}} & \qw & \qw & \qw & \gate{\mathrm{T}} & \qw & \qw & \ctrl{-2} & \qw & \qw & \ctrl{-3} & \qw & \qw & \qw & \ctrl{-1} & \qw & \qw & \qw & \ctrl{-6} & \qw & \qw & \qw\\
\\ }}

%% file: fig1b.tex
\begin{tikzpicture}
  \definecolor{qpu0}{HTML}{1f77b4}
  \definecolor{qpu1}{HTML}{ff7f0e}
  \tikzset{qubit/.style={circle, draw=black, minimum size=13pt, inner sep=0pt, align=center}, gate/.style={black, -}}
  \def\radius{26.0pt}
  \def\labeloffset{52.0pt}
  \def\xspacing{78.0pt}
  \begin{scope}[xshift=0.0pt, yshift=0pt]{
    \node[name=t0_label] at (0, \labeloffset) {$t=0$};
    \node[qubit, fill=qpu0, text=white, name=q00] at ({90}:\radius) {$q_0$};
    \node[qubit, fill=qpu1, text=white, name=q01] at ({45}:\radius) {$q_1$};
    \node[qubit, fill=qpu0, text=white, name=q02] at ({0}:\radius) {$q_2$};
    \node[qubit, fill=qpu1, text=white, name=q03] at ({-45}:\radius) {$q_3$};
    \node[qubit, fill=qpu1, text=white, name=q04] at ({-90}:\radius) {$q_4$};
    \node[qubit, fill=qpu0, text=white, name=q05] at ({-135}:\radius) {$q_5$};
    \node[qubit, fill=qpu1, text=white, name=q06] at ({-180}:\radius) {$q_6$};
    \node[qubit, fill=qpu0, text=white, name=q07] at ({-225}:\radius) {$q_7$};
    \begin{scope}[gate]
      \draw (q03) -- (q06);
      \draw (q05) -- (q07);
    \end{scope}
  }\end{scope}
  \begin{scope}[xshift=78.0pt, yshift=0pt]{
    \node[name=t1_label] at (0, \labeloffset) {$t=1$};
    \node[qubit, fill=qpu0, text=white, name=q10] at ({90}:\radius) {$q_0$};
    \node[qubit, fill=qpu1, text=white, name=q11] at ({45}:\radius) {$q_1$};
    \node[qubit, fill=qpu0, text=white, name=q12] at ({0}:\radius) {$q_2$};
    \node[qubit, fill=qpu1, text=white, name=q13] at ({-45}:\radius) {$q_3$};
    \node[qubit, fill=qpu1, text=white, name=q14] at ({-90}:\radius) {$q_4$};
    \node[qubit, fill=qpu0, text=white, name=q15] at ({-135}:\radius) {$q_5$};
    \node[qubit, fill=qpu1, text=white, name=q16] at ({-180}:\radius) {$q_6$};
    \node[qubit, fill=qpu0, text=white, name=q17] at ({-225}:\radius) {$q_7$};
    \begin{scope}[gate]
      \draw (q13) -- (q14);
      \draw (q15) -- (q17);
    \end{scope}
  }\end{scope}
  \begin{scope}[xshift=156.0pt, yshift=0pt]{
    \node[name=t2_label] at (0, \labeloffset) {$t=2$};
    \node[qubit, fill=qpu0, text=white, name=q20] at ({90}:\radius) {$q_0$};
    \node[qubit, fill=qpu1, text=white, name=q21] at ({45}:\radius) {$q_1$};
    \node[qubit, fill=qpu0, text=white, name=q22] at ({0}:\radius) {$q_2$};
    \node[qubit, fill=qpu1, text=white, name=q23] at ({-45}:\radius) {$q_3$};
    \node[qubit, fill=qpu1, text=white, name=q24] at ({-90}:\radius) {$q_4$};
    \node[qubit, fill=qpu0, text=white, name=q25] at ({-135}:\radius) {$q_5$};
    \node[qubit, fill=qpu1, text=white, name=q26] at ({-180}:\radius) {$q_6$};
    \node[qubit, fill=qpu0, text=white, name=q27] at ({-225}:\radius) {$q_7$};
    \begin{scope}[gate]
      \draw (q20) -- (q25);
      \draw (q22) -- (q24);
      \draw (q23) -- (q26);
    \end{scope}
  }\end{scope}
  \begin{scope}[xshift=234.0pt, yshift=0pt]{
    \node[name=t3_label] at (0, \labeloffset) {$t=3$};
    \node[qubit, fill=qpu0, text=white, name=q30] at ({90}:\radius) {$q_0$};
    \node[qubit, fill=qpu1, text=white, name=q31] at ({45}:\radius) {$q_1$};
    \node[qubit, fill=qpu0, text=white, name=q32] at ({0}:\radius) {$q_2$};
    \node[qubit, fill=qpu1, text=white, name=q33] at ({-45}:\radius) {$q_3$};
    \node[qubit, fill=qpu1, text=white, name=q34] at ({-90}:\radius) {$q_4$};
    \node[qubit, fill=qpu0, text=white, name=q35] at ({-135}:\radius) {$q_5$};
    \node[qubit, fill=qpu1, text=white, name=q36] at ({-180}:\radius) {$q_6$};
    \node[qubit, fill=qpu0, text=white, name=q37] at ({-225}:\radius) {$q_7$};
    \begin{scope}[gate]
      \draw (q32) -- (q35);
      \draw (q34) -- (q36);
    \end{scope}
  }\end{scope}
  \begin{scope}[xshift=312.0pt, yshift=0pt]{
    \node[name=t4_label] at (0, \labeloffset) {$t=4$};
    \node[qubit, fill=qpu0, text=white, name=q40] at ({90}:\radius) {$q_0$};
    \node[qubit, fill=qpu1, text=white, name=q41] at ({45}:\radius) {$q_1$};
    \node[qubit, fill=qpu0, text=white, name=q42] at ({0}:\radius) {$q_2$};
    \node[qubit, fill=qpu1, text=white, name=q43] at ({-45}:\radius) {$q_3$};
    \node[qubit, fill=qpu1, text=white, name=q44] at ({-90}:\radius) {$q_4$};
    \node[qubit, fill=qpu0, text=white, name=q45] at ({-135}:\radius) {$q_5$};
    \node[qubit, fill=qpu1, text=white, name=q46] at ({-180}:\radius) {$q_6$};
    \node[qubit, fill=qpu0, text=white, name=q47] at ({-225}:\radius) {$q_7$};
    \begin{scope}[gate]
      \draw (q40) -- (q43);
      \draw (q42) -- (q46);
      \draw (q45) -- (q47);
    \end{scope}
  }\end{scope}
  \begin{scope}[xshift=390.0pt, yshift=0pt]{
    \node[name=t5_label] at (0, \labeloffset) {$t=5$};
    \node[qubit, fill=qpu0, text=white, name=q50] at ({90}:\radius) {$q_0$};
    \node[qubit, fill=qpu1, text=white, name=q51] at ({45}:\radius) {$q_1$};
    \node[qubit, fill=qpu0, text=white, name=q52] at ({0}:\radius) {$q_2$};
    \node[qubit, fill=qpu0, text=white, name=q53] at ({-45}:\radius) {$q_3$};
    \node[qubit, fill=qpu1, text=white, name=q54] at ({-90}:\radius) {$q_4$};
    \node[qubit, fill=qpu0, text=white, name=q55] at ({-135}:\radius) {$q_5$};
    \node[qubit, fill=qpu1, text=white, name=q56] at ({-180}:\radius) {$q_6$};
    \node[qubit, fill=qpu1, text=white, name=q57] at ({-225}:\radius) {$q_7$};
    \begin{scope}[gate]
      \draw (q51) -- (q56);
      \draw (q53) -- (q55);
      \draw (q54) -- (q57);
    \end{scope}
  }\end{scope}
  \begin{scope}[xshift=468.0pt, yshift=0pt]{
    \node[name=t6_label] at (0, \labeloffset) {$t=6$};
    \node[qubit, fill=qpu0, text=white, name=q60] at ({90}:\radius) {$q_0$};
    \node[qubit, fill=qpu1, text=white, name=q61] at ({45}:\radius) {$q_1$};
    \node[qubit, fill=qpu0, text=white, name=q62] at ({0}:\radius) {$q_2$};
    \node[qubit, fill=qpu0, text=white, name=q63] at ({-45}:\radius) {$q_3$};
    \node[qubit, fill=qpu1, text=white, name=q64] at ({-90}:\radius) {$q_4$};
    \node[qubit, fill=qpu0, text=white, name=q65] at ({-135}:\radius) {$q_5$};
    \node[qubit, fill=qpu1, text=white, name=q66] at ({-180}:\radius) {$q_6$};
    \node[qubit, fill=qpu1, text=white, name=q67] at ({-225}:\radius) {$q_7$};
    \begin{scope}[gate]
      \draw (q60) -- (q62);
      \draw (q66) -- (q67);
    \end{scope}
  }\end{scope}
  \begin{scope}[xshift=546.0pt, yshift=0pt]{
    \node[name=t7_label] at (0, \labeloffset) {$t=7$};
    \node[qubit, fill=qpu0, text=white, name=q70] at ({90}:\radius) {$q_0$};
    \node[qubit, fill=qpu1, text=white, name=q71] at ({45}:\radius) {$q_1$};
    \node[qubit, fill=qpu0, text=white, name=q72] at ({0}:\radius) {$q_2$};
    \node[qubit, fill=qpu0, text=white, name=q73] at ({-45}:\radius) {$q_3$};
    \node[qubit, fill=qpu1, text=white, name=q74] at ({-90}:\radius) {$q_4$};
    \node[qubit, fill=qpu0, text=white, name=q75] at ({-135}:\radius) {$q_5$};
    \node[qubit, fill=qpu1, text=white, name=q76] at ({-180}:\radius) {$q_6$};
    \node[qubit, fill=qpu1, text=white, name=q77] at ({-225}:\radius) {$q_7$};
    \begin{scope}[gate]
      \draw (q71) -- (q77);
      \draw (q73) -- (q75);
      \draw (q74) -- (q76);
    \end{scope}
  }\end{scope}
\end{tikzpicture}

%% file: fig2a.tex
\begingroup
  \makeatletter
  \providecommand\color[2][]{%
    \GenericError{(gnuplot) \space\space\space\@spaces}{%
      Package color not loaded in conjunction with
      terminal option `colourtext'%
    }{See the gnuplot documentation for explanation.%
    }{Either use 'blacktext' in gnuplot or load the package
      color.sty in LaTeX.}%
    \renewcommand\color[2][]{}%
  }%
  \providecommand\includegraphics[2][]{%
    \GenericError{(gnuplot) \space\space\space\@spaces}{%
      Package graphicx or graphics not loaded%
    }{See the gnuplot documentation for explanation.%
    }{The gnuplot epslatex terminal needs graphicx.sty or graphics.sty.}%
    \renewcommand\includegraphics[2][]{}%
  }%
  \providecommand\rotatebox[2]{#2}%
  \@ifundefined{ifGPcolor}{%
    \newif\ifGPcolor
    \GPcolortrue
  }{}%
  \@ifundefined{ifGPblacktext}{%
    \newif\ifGPblacktext
    \GPblacktexttrue
  }{}%
  \let\gplgaddtomacro\g@addto@macro
  \gdef\gplbacktext{}%
  \gdef\gplfronttext{}%
  \makeatother
  \ifGPblacktext
    \def\colorrgb#1{}%
    \def\colorgray#1{}%
  \else
    \ifGPcolor
      \def\colorrgb#1{\color[rgb]{#1}}%
      \def\colorgray#1{\color[gray]{#1}}%
      \expandafter\def\csname LTw\endcsname{\color{white}}%
      \expandafter\def\csname LTb\endcsname{\color{black}}%
      \expandafter\def\csname LTa\endcsname{\color{black}}%
      \expandafter\def\csname LT0\endcsname{\color[rgb]{1,0,0}}%
      \expandafter\def\csname LT1\endcsname{\color[rgb]{0,1,0}}%
      \expandafter\def\csname LT2\endcsname{\color[rgb]{0,0,1}}%
      \expandafter\def\csname LT3\endcsname{\color[rgb]{1,0,1}}%
      \expandafter\def\csname LT4\endcsname{\color[rgb]{0,1,1}}%
      \expandafter\def\csname LT5\endcsname{\color[rgb]{1,1,0}}%
      \expandafter\def\csname LT6\endcsname{\color[rgb]{0,0,0}}%
      \expandafter\def\csname LT7\endcsname{\color[rgb]{1,0.3,0}}%
      \expandafter\def\csname LT8\endcsname{\color[rgb]{0.5,0.5,0.5}}%
    \else
      \def\colorrgb#1{\color{black}}%
      \def\colorgray#1{\color[gray]{#1}}%
      \expandafter\def\csname LTw\endcsname{\color{white}}%
      \expandafter\def\csname LTb\endcsname{\color{black}}%
      \expandafter\def\csname LTa\endcsname{\color{black}}%
      \expandafter\def\csname LT0\endcsname{\color{black}}%
      \expandafter\def\csname LT1\endcsname{\color{black}}%
      \expandafter\def\csname LT2\endcsname{\color{black}}%
      \expandafter\def\csname LT3\endcsname{\color{black}}%
      \expandafter\def\csname LT4\endcsname{\color{black}}%
      \expandafter\def\csname LT5\endcsname{\color{black}}%
      \expandafter\def\csname LT6\endcsname{\color{black}}%
      \expandafter\def\csname LT7\endcsname{\color{black}}%
      \expandafter\def\csname LT8\endcsname{\color{black}}%
    \fi
  \fi
    \setlength{\unitlength}{0.0500bp}%
    \ifx\gptboxheight\undefined%
      \newlength{\gptboxheight}%
      \newlength{\gptboxwidth}%
      \newsavebox{\gptboxtext}%
    \fi%
    \setlength{\fboxrule}{0.5pt}%
    \setlength{\fboxsep}{1pt}%
    \definecolor{tbcol}{rgb}{1,1,1}%
\begin{picture}(7200.00,4320.00)%
    \gplgaddtomacro\gplbacktext{%
      \csname LTb\endcsname
      \put(714,633){\makebox(0,0)[r]{\strut{}$0$}}%
      \csname LTb\endcsname
      \put(714,1214){\makebox(0,0)[r]{\strut{}$200$}}%
      \csname LTb\endcsname
      \put(714,1796){\makebox(0,0)[r]{\strut{}$400$}}%
      \csname LTb\endcsname
      \put(714,2378){\makebox(0,0)[r]{\strut{}$600$}}%
      \csname LTb\endcsname
      \put(714,2960){\makebox(0,0)[r]{\strut{}$800$}}%
      \csname LTb\endcsname
      \put(714,3542){\makebox(0,0)[r]{\strut{}$1000$}}%
      \csname LTb\endcsname
      \put(714,4124){\makebox(0,0)[r]{\strut{}$1200$}}%
      \csname LTb\endcsname
      \put(2027,386){\makebox(0,0){\strut{}8}}%
      \csname LTb\endcsname
      \put(3241,386){\makebox(0,0){\strut{}16}}%
      \csname LTb\endcsname
      \put(4456,386){\makebox(0,0){\strut{}32}}%
      \csname LTb\endcsname
      \put(5671,386){\makebox(0,0){\strut{}64}}%
    }%
    \gplgaddtomacro\gplfronttext{%
      \csname LTb\endcsname
      \put(1987,3965){\makebox(0,0)[r]{\strut{}METIS}}%
      \csname LTb\endcsname
      \put(1987,3789){\makebox(0,0)[r]{\strut{}Beam search}}%
      \csname LTb\endcsname
      \put(2270,1114){\makebox(0,0){\strut{}27.5\%}}%
      \csname LTb\endcsname
      \put(3484,1365){\makebox(0,0){\strut{}32.3\%}}%
      \csname LTb\endcsname
      \put(4699,2061){\makebox(0,0){\strut{}24.9\%}}%
      \csname LTb\endcsname
      \put(5914,3670){\makebox(0,0){\strut{}15.1\%}}%
      \csname LTb\endcsname
      \put(161,2378){\rotatebox{-270.00}{\makebox(0,0){\strut{}Communication cost $C$}}}%
      \csname LTb\endcsname
      \put(3849,123){\makebox(0,0){\strut{}Number of qubits $N$}}%
    }%
    \gplbacktext
    \put(0,0){\includegraphics[width={360.00bp},height={216.00bp}]{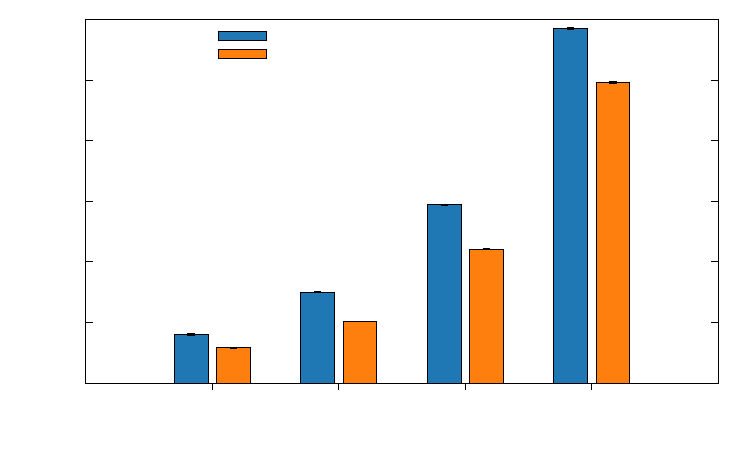}}%
    \gplfronttext
  \end{picture}%
\endgroup

%% file: fig2b.tex
\begingroup
  \makeatletter
  \providecommand\color[2][]{%
    \GenericError{(gnuplot) \space\space\space\@spaces}{%
      Package color not loaded in conjunction with
      terminal option `colourtext'%
    }{See the gnuplot documentation for explanation.%
    }{Either use 'blacktext' in gnuplot or load the package
      color.sty in LaTeX.}%
    \renewcommand\color[2][]{}%
  }%
  \providecommand\includegraphics[2][]{%
    \GenericError{(gnuplot) \space\space\space\@spaces}{%
      Package graphicx or graphics not loaded%
    }{See the gnuplot documentation for explanation.%
    }{The gnuplot epslatex terminal needs graphicx.sty or graphics.sty.}%
    \renewcommand\includegraphics[2][]{}%
  }%
  \providecommand\rotatebox[2]{#2}%
  \@ifundefined{ifGPcolor}{%
    \newif\ifGPcolor
    \GPcolortrue
  }{}%
  \@ifundefined{ifGPblacktext}{%
    \newif\ifGPblacktext
    \GPblacktexttrue
  }{}%
  \let\gplgaddtomacro\g@addto@macro
  \gdef\gplbacktext{}%
  \gdef\gplfronttext{}%
  \makeatother
  \ifGPblacktext
    \def\colorrgb#1{}%
    \def\colorgray#1{}%
  \else
    \ifGPcolor
      \def\colorrgb#1{\color[rgb]{#1}}%
      \def\colorgray#1{\color[gray]{#1}}%
      \expandafter\def\csname LTw\endcsname{\color{white}}%
      \expandafter\def\csname LTb\endcsname{\color{black}}%
      \expandafter\def\csname LTa\endcsname{\color{black}}%
      \expandafter\def\csname LT0\endcsname{\color[rgb]{1,0,0}}%
      \expandafter\def\csname LT1\endcsname{\color[rgb]{0,1,0}}%
      \expandafter\def\csname LT2\endcsname{\color[rgb]{0,0,1}}%
      \expandafter\def\csname LT3\endcsname{\color[rgb]{1,0,1}}%
      \expandafter\def\csname LT4\endcsname{\color[rgb]{0,1,1}}%
      \expandafter\def\csname LT5\endcsname{\color[rgb]{1,1,0}}%
      \expandafter\def\csname LT6\endcsname{\color[rgb]{0,0,0}}%
      \expandafter\def\csname LT7\endcsname{\color[rgb]{1,0.3,0}}%
      \expandafter\def\csname LT8\endcsname{\color[rgb]{0.5,0.5,0.5}}%
    \else
      \def\colorrgb#1{\color{black}}%
      \def\colorgray#1{\color[gray]{#1}}%
      \expandafter\def\csname LTw\endcsname{\color{white}}%
      \expandafter\def\csname LTb\endcsname{\color{black}}%
      \expandafter\def\csname LTa\endcsname{\color{black}}%
      \expandafter\def\csname LT0\endcsname{\color{black}}%
      \expandafter\def\csname LT1\endcsname{\color{black}}%
      \expandafter\def\csname LT2\endcsname{\color{black}}%
      \expandafter\def\csname LT3\endcsname{\color{black}}%
      \expandafter\def\csname LT4\endcsname{\color{black}}%
      \expandafter\def\csname LT5\endcsname{\color{black}}%
      \expandafter\def\csname LT6\endcsname{\color{black}}%
      \expandafter\def\csname LT7\endcsname{\color{black}}%
      \expandafter\def\csname LT8\endcsname{\color{black}}%
    \fi
  \fi
    \setlength{\unitlength}{0.0500bp}%
    \ifx\gptboxheight\undefined%
      \newlength{\gptboxheight}%
      \newlength{\gptboxwidth}%
      \newsavebox{\gptboxtext}%
    \fi%
    \setlength{\fboxrule}{0.5pt}%
    \setlength{\fboxsep}{1pt}%
    \definecolor{tbcol}{rgb}{1,1,1}%
\begin{picture}(7200.00,4320.00)%
    \gplgaddtomacro\gplbacktext{%
      \csname LTb\endcsname
      \put(714,633){\makebox(0,0)[r]{\strut{}$0$}}%
      \csname LTb\endcsname
      \put(714,1131){\makebox(0,0)[r]{\strut{}$200$}}%
      \csname LTb\endcsname
      \put(714,1630){\makebox(0,0)[r]{\strut{}$400$}}%
      \csname LTb\endcsname
      \put(714,2129){\makebox(0,0)[r]{\strut{}$600$}}%
      \csname LTb\endcsname
      \put(714,2627){\makebox(0,0)[r]{\strut{}$800$}}%
      \csname LTb\endcsname
      \put(714,3126){\makebox(0,0)[r]{\strut{}$1000$}}%
      \csname LTb\endcsname
      \put(714,3625){\makebox(0,0)[r]{\strut{}$1200$}}%
      \csname LTb\endcsname
      \put(714,4124){\makebox(0,0)[r]{\strut{}$1400$}}%
      \csname LTb\endcsname
      \put(2027,386){\makebox(0,0){\strut{}32}}%
      \csname LTb\endcsname
      \put(3241,386){\makebox(0,0){\strut{}64}}%
      \csname LTb\endcsname
      \put(4456,386){\makebox(0,0){\strut{}128}}%
      \csname LTb\endcsname
      \put(5671,386){\makebox(0,0){\strut{}256}}%
    }%
    \gplgaddtomacro\gplfronttext{%
      \csname LTb\endcsname
      \put(1987,3965){\makebox(0,0)[r]{\strut{}METIS}}%
      \csname LTb\endcsname
      \put(1987,3789){\makebox(0,0)[r]{\strut{}Beam search}}%
      \csname LTb\endcsname
      \put(2270,1044){\makebox(0,0){\strut{}12.3\%}}%
      \csname LTb\endcsname
      \put(3484,1316){\makebox(0,0){\strut{}21.5\%}}%
      \csname LTb\endcsname
      \put(4699,1877){\makebox(0,0){\strut{}24.9\%}}%
      \csname LTb\endcsname
      \put(5914,2977){\makebox(0,0){\strut{}28.5\%}}%
      \csname LTb\endcsname
      \put(161,2378){\rotatebox{-270.00}{\makebox(0,0){\strut{}Communication cost $C$}}}%
      \csname LTb\endcsname
      \put(3849,123){\makebox(0,0){\strut{}Depth $T$}}%
    }%
    \gplbacktext
    \put(0,0){\includegraphics[width={360.00bp},height={216.00bp}]{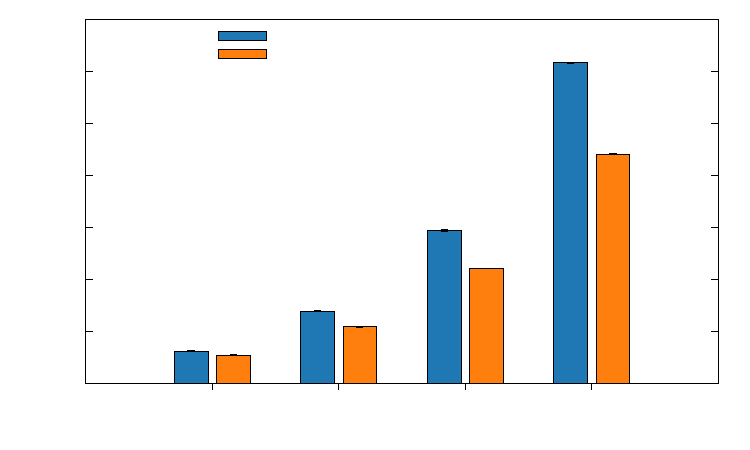}}%
    \gplfronttext
  \end{picture}%
\endgroup

%% file: fig3.tex
\begingroup
  \makeatletter
  \providecommand\color[2][]{%
    \GenericError{(gnuplot) \space\space\space\@spaces}{%
      Package color not loaded in conjunction with
      terminal option `colourtext'%
    }{See the gnuplot documentation for explanation.%
    }{Either use 'blacktext' in gnuplot or load the package
      color.sty in LaTeX.}%
    \renewcommand\color[2][]{}%
  }%
  \providecommand\includegraphics[2][]{%
    \GenericError{(gnuplot) \space\space\space\@spaces}{%
      Package graphicx or graphics not loaded%
    }{See the gnuplot documentation for explanation.%
    }{The gnuplot epslatex terminal needs graphicx.sty or graphics.sty.}%
    \renewcommand\includegraphics[2][]{}%
  }%
  \providecommand\rotatebox[2]{#2}%
  \@ifundefined{ifGPcolor}{%
    \newif\ifGPcolor
    \GPcolortrue
  }{}%
  \@ifundefined{ifGPblacktext}{%
    \newif\ifGPblacktext
    \GPblacktexttrue
  }{}%
  \let\gplgaddtomacro\g@addto@macro
  \gdef\gplbacktext{}%
  \gdef\gplfronttext{}%
  \makeatother
  \ifGPblacktext
    \def\colorrgb#1{}%
    \def\colorgray#1{}%
  \else
    \ifGPcolor
      \def\colorrgb#1{\color[rgb]{#1}}%
      \def\colorgray#1{\color[gray]{#1}}%
      \expandafter\def\csname LTw\endcsname{\color{white}}%
      \expandafter\def\csname LTb\endcsname{\color{black}}%
      \expandafter\def\csname LTa\endcsname{\color{black}}%
      \expandafter\def\csname LT0\endcsname{\color[rgb]{1,0,0}}%
      \expandafter\def\csname LT1\endcsname{\color[rgb]{0,1,0}}%
      \expandafter\def\csname LT2\endcsname{\color[rgb]{0,0,1}}%
      \expandafter\def\csname LT3\endcsname{\color[rgb]{1,0,1}}%
      \expandafter\def\csname LT4\endcsname{\color[rgb]{0,1,1}}%
      \expandafter\def\csname LT5\endcsname{\color[rgb]{1,1,0}}%
      \expandafter\def\csname LT6\endcsname{\color[rgb]{0,0,0}}%
      \expandafter\def\csname LT7\endcsname{\color[rgb]{1,0.3,0}}%
      \expandafter\def\csname LT8\endcsname{\color[rgb]{0.5,0.5,0.5}}%
    \else
      \def\colorrgb#1{\color{black}}%
      \def\colorgray#1{\color[gray]{#1}}%
      \expandafter\def\csname LTw\endcsname{\color{white}}%
      \expandafter\def\csname LTb\endcsname{\color{black}}%
      \expandafter\def\csname LTa\endcsname{\color{black}}%
      \expandafter\def\csname LT0\endcsname{\color{black}}%
      \expandafter\def\csname LT1\endcsname{\color{black}}%
      \expandafter\def\csname LT2\endcsname{\color{black}}%
      \expandafter\def\csname LT3\endcsname{\color{black}}%
      \expandafter\def\csname LT4\endcsname{\color{black}}%
      \expandafter\def\csname LT5\endcsname{\color{black}}%
      \expandafter\def\csname LT6\endcsname{\color{black}}%
      \expandafter\def\csname LT7\endcsname{\color{black}}%
      \expandafter\def\csname LT8\endcsname{\color{black}}%
    \fi
  \fi
    \setlength{\unitlength}{0.0500bp}%
    \ifx\gptboxheight\undefined%
      \newlength{\gptboxheight}%
      \newlength{\gptboxwidth}%
      \newsavebox{\gptboxtext}%
    \fi%
    \setlength{\fboxrule}{0.5pt}%
    \setlength{\fboxsep}{1pt}%
    \definecolor{tbcol}{rgb}{1,1,1}%
\begin{picture}(7200.00,4320.00)%
    \gplgaddtomacro\gplbacktext{%
      \csname LTb\endcsname
      \put(714,633){\makebox(0,0)[r]{\strut{}$800$}}%
      \csname LTb\endcsname
      \put(714,1069){\makebox(0,0)[r]{\strut{}$900$}}%
      \csname LTb\endcsname
      \put(714,1505){\makebox(0,0)[r]{\strut{}$1000$}}%
      \csname LTb\endcsname
      \put(714,1942){\makebox(0,0)[r]{\strut{}$1100$}}%
      \csname LTb\endcsname
      \put(714,2378){\makebox(0,0)[r]{\strut{}$1200$}}%
      \csname LTb\endcsname
      \put(714,2814){\makebox(0,0)[r]{\strut{}$1300$}}%
      \csname LTb\endcsname
      \put(714,3251){\makebox(0,0)[r]{\strut{}$1400$}}%
      \csname LTb\endcsname
      \put(714,3687){\makebox(0,0)[r]{\strut{}$1500$}}%
      \csname LTb\endcsname
      \put(714,4124){\makebox(0,0)[r]{\strut{}$1600$}}%
      \csname LTb\endcsname
      \put(2027,386){\makebox(0,0){\strut{}Complete}}%
      \csname LTb\endcsname
      \put(3241,386){\makebox(0,0){\strut{}Cycle}}%
      \csname LTb\endcsname
      \put(4456,386){\makebox(0,0){\strut{}Star}}%
      \csname LTb\endcsname
      \put(5671,386){\makebox(0,0){\strut{}Path}}%
    }%
    \gplgaddtomacro\gplfronttext{%
      \csname LTb\endcsname
      \put(1987,3965){\makebox(0,0)[r]{\strut{}METIS}}%
      \csname LTb\endcsname
      \put(1987,3789){\makebox(0,0)[r]{\strut{}Beam search}}%
      \csname LTb\endcsname
      \put(2270,965){\makebox(0,0){\strut{}8.1\%}}%
      \csname LTb\endcsname
      \put(3484,1854){\makebox(0,0){\strut{}13.7\%}}%
      \csname LTb\endcsname
      \put(4699,2628){\makebox(0,0){\strut{}10.3\%}}%
      \csname LTb\endcsname
      \put(5914,3063){\makebox(0,0){\strut{}13.1\%}}%
      \csname LTb\endcsname
      \put(161,2378){\rotatebox{-270.00}{\makebox(0,0){\strut{}Communication cost $C$}}}%
      \csname LTb\endcsname
      \put(3849,123){\makebox(0,0){\strut{}Network topology (graph)}}%
    }%
    \gplbacktext
    \put(0,0){\includegraphics[width={360.00bp},height={216.00bp}]{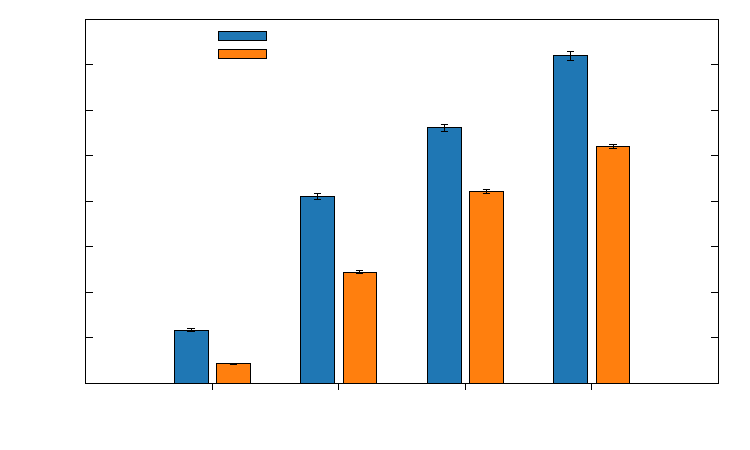}}%
    \gplfronttext
  \end{picture}%
\endgroup